\documentclass{ws-ijmpa}

\newcommand{\jp}{J/\psi}
\newcommand{\jpsi}{J/\psi}
\newcommand{\psip}{\psi^{'}}

\newcommand{\pipi}{\pi^{+}\pi^{-}}

\newcommand{\rt}{\rightarrow}
\newcommand{\etal}{\em et al.}

\begin{document}

\markboth{Stephen L. Olsen}
{The X(3872)}

\catchline{}{}{}{}{}
%
%%%%%%%%%%%%%%%%%%%%%%%%%%%%%%%%%%%%%%%%%%%%%%%%%%%%%%%%%%%%%%%%%%%%

\title{SEARCH FOR A CHARMONIUM ASSIGNMENT FOR THE X(3872)}

\author{\footnotesize STEPHEN L. OLSEN\footnote{
email:solsen@phys.hawaii.edu}\\
Representing the Belle Collaboration}

%\author{\footnotesize STEPHEN L. OLSEN\footnote{Representing the
%Belle Collaboration.~~~email:solsen@phys.hawaii.edu} }

\address{Department of Physics and Astronomy, University of Hawaii at 
Manoa\\ 
2505 Correa Road,  Honolulu, Hawaii 96822,
U.S.A.}

\maketitle

%\pub{Received (Day Month Year)}{Revised (Day Month Year)}

\begin{abstract}
We report recent results on the properties of the $X(3872)$ produced
via the $B^{+} \rt K^{+} X(3872)$ decay process in the Belle detector. 
We compare these properties with expectations for possible  
charmonium-state assignments.

\end{abstract}

\section{Introduction}	%) A SECTION HEADING

\begin{figure}
\centerline{\psfig{file=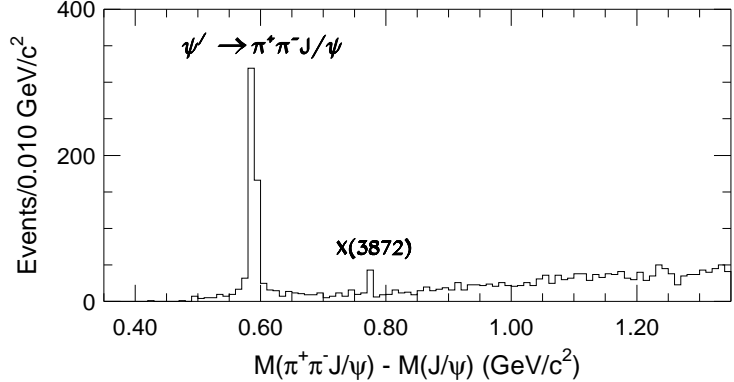,width=10cm}}
\vspace*{8pt}
\caption{$M_{\pipi\jpsi}-M_{\jpsi}$ for $B\rt\pipi\jpsi$ decays
seen in the Belle experiment.  The large peak at 0.59~GeV 
corresponds to $B\rt K\psip$; $\psip\rt\pipi\jpsi$ events.
The peak at 0.776~GeV is the signal for $X(3872)\rt\pipi\jpsi$.
\label{fig:pipijpsi}}
\end{figure}

The $X(3872)$ is a narrow state that decays into $\pipi\jpsi$.  Although its 
mass,\footnote{This mass value is a weighted
 average of the results from Refs.~[1-4].}
$M_X = 3871.9\pm 0.5$~MeV, is well above the $D\bar{D}$
open-charm 
threshold, its width is narrow; the current experimental upper 
limit on its width is $\Gamma < 2.3$~MeV~(90\%~CL).\cite{skchoi_1}  It 
was first seen in exclusive $B\rt K X(3872)$ decays by the Belle 
experiment\cite{skchoi_1} (see Fig.~\ref{fig:pipijpsi}) and
subsequently seen in inclusive $p\bar{p}$ collisons by CDF\cite{CDF} and
D0.\cite{D0}  Recently, the BaBar group confirmed its production in
exclusive $B$-meson decays.\cite{BaBar_1}   Although it looks like
a typical charmonium particle ($i.e.$ a $c\bar{c}$ meson),  its
assignment to any of the as-yet unseen narrow charmonium states
has proven to be problematic.  

In this report I describe recent results on the $X(3872)$
and compare its known properties with expectations for 
possible charmonium assignments.  

\section{Properties of the $X(3872)$ }

First I summarize the known properties of the $X(3872)$:
\begin{romanlist}
\item
It decays to $\pipi\jpsi$.
\item
Its mass is very close to the $M_{D^0} + M_{D^{*0}}$ mass
threshold.
\item
It is narrow ($\Gamma < 2.3$~MeV).
\item
Although its mass is more that 140~MeV above the $D\bar{D}$ mass
threshold, decays to 
$D\bar{D}$ are not seen;  Belle\cite{chistov} reports
$\Gamma(X\rt D\bar{D})/
\Gamma(X\rt\pipi\jpsi)<7~~{\rm (90\%~CL)}.$
(The same ratio for the $\psi(3770)$, which
is only about 30~MeV above the $D\bar{D}$ threshold,
is\cite{BES}  $> 160$.)
The absence of $D\bar{D}$ decays, taken together with
fact that it is narrow 
indicates that $D\bar{D}$ final states are probably not allowed.  This
suggests that the natural quantum number sequence 
$J^P = 0^+ , 1^-, 2^+,$~etc. is ruled out.
\item
In the $\pipi\jpsi$ decays, the dipion masses 
tend to concentrate near the mass of the $\rho(770)$ meson. 
The decay of a $c\bar{c}$ charmonium state to $\rho\jpsi$
would violate isospin and isospin-violating charmonium transitions
are strongly suppressed.
If the dipions are in fact coming from $\rho\rt\pipi$ decays,
the charge-conjugation parity of the $X(3872)$ would be
$C=+1$ and $X\rt\pi^0\pi^0\jpsi$ decays would be forbidden.
Otherwise, the $X(3872)$ would have $C=-1$ and 
$\Gamma(X\rt\pi^0\pi^0\jpsi)\simeq \frac{1}{2} \Gamma(X\rt\pipi\jpsi)$. 
\item
The decay $X(3872)\rt \gamma \chi_{c1}$ is not seen:\cite{skchoi_1}
$$\Gamma(X\rt \gamma\chi_{c1})/
\Gamma(X\rt\pipi\jpsi)<0.89~~{\rm (90\%~CL)}.$$
\item
It is seen in exclusive $B\rt KX$ decays with the product
branching fraction\cite{skchoi_1,BaBar_1}
\begin{equation}
{\cal B}(B^-\rt K^- X)\times
{\cal B}(X\rt\pipi\jpsi) = (1.3 \pm 0.3)\times 10^{-5}.
\label{eq:product_br}
\end{equation}
This suggests that high values of $J$ are not likely (see Sect.~4.3 
below).
\end{romanlist}

\section{Charmonium Possibilities}

%\begin{figure}
%\centerline{\psfig{file=charmonium.ps,width=7.5cm}}
%\vspace*{8pt}
%\caption{The charmonium level diagram.  The horizontal dashed
%lines indicate the $D\bar{D}$ and $D\bar{D^*}$ open charm
%thresholds.  The masses of the as-yet
%undiscovered states are taken from Ref.~[7].
%\label{fig:charmonium}}
%\end{figure}

%The charmonium level diagram is shown in 
%Fig.~\ref{fig:charmonium}.  Here the mass values
%are taken from Ref.~[7].
We consider  as-yet unseen charmonium states 
with expected masses within $\sim$200~MeV of $3872$~MeV, 
with unnatural quantum numbers ($i.e.~J ^P = 0^-, 1^+, 2^-$, etc.),
and with spin angular momentum $J<3$. 
There are five candidate states that meet these
criteria: the 
$1^3D_2$, $2^1P_1$, $2 ^3P_1$, $1 ^1D_1$, and $3^1S_0$.
We also consider the $1^3D_3$, even though
it fails two of our criteria --- it has $J=3$ and 
decays to $D\bar{D}$ are allowed.
This has been promoted to the candidate list
because some authors\cite{barnes,elq_2} have 
identified this as a candidate based primarily
on the observation that $\psi_3 \rt D\bar{D}$ decays  
are suppressed
by an $L=3$ angular momentum barrier.

\begin{table}[h]
\tbl{\label{tbl:charmonium}Some properties of the candidate charmonium states.}
{\begin{tabular}{@{}ccccc@{}} \toprule
State     & nickname     & $J^{PC}$ & $M_{predicted}$ (MeV)  &  $\Gamma_{predicted}$ (MeV)  \\ \colrule
%         &              &          &    (MeV)               &         (MeV)                \\ \colrule
$1^3D_2$  & $\psi_2$     & $2^{--}$ &    3838                &         $0.7$                \\
$2^1P_1$  &  $h_c'$      & $1^{+-}$ &    3953                &         $1.6$                \\
$1^3D_3$  & $\psi_3$     & $3^{--}$ &    3849                &         $4.8$                \\
$2^3P_1$  & $\chi_{c1}'$ & $1^{++}$ &    3956                &         $1.7$                \\
$1^1D_2$  & $\eta_{c2}$  & $2^{-+}$ &    3837                &         $0.9$                \\
$3^1S_0$  & $\eta_c''$   & $0^{-+}$ &    4060                &         $\sim 20$            \\ \botrule

\end{tabular}}
\end{table}

The six candidate states are summarized in Table~\ref{tbl:charmonium}
roughly in the order of their plausibility.  We include in the Table
the quantum numbers and a potential model
prediction\cite{godfrey-isgur} 
for the mass, and total width values\footnote{The predicted
width for the $\eta_c''$ is taken to be the same as the 
(poorly known) $\eta_c$ width.}
that are from Ref.~[8] and computed using a 3872~MeV mass value.

In the following I discuss each candidate assignment one-by-one
in the context of measurements in progress that are intended
to confirm or disallow that assignment.

\section{\boldmath $C=-1$ assignments}

If the $X(3872)$ is a $C=-1$ state, the $\pipi\jpsi$ transition is 
isospin conserving and not suppressed.   Thus, one of these assignments
would seem to be more reasonable.  For this case,\cite{voloshin} the $X\rt \pi^0\pi^0\jpsi$
partial decay width would be about half of that for $\pipi\jpsi$.
The $\pi^0\pi^0\jpsi$ channel is more experimentally challenging  
than $\pipi\jpsi$ and there have been no  results 
reported to date.  Belle hopes to report a measurement
of this channel in Summer 2004.

\subsection{\boldmath $X(3872) = \psi_2$?}
In the charmonium model there are two states expected to 
have mass between $2M_D$ and $M_D + M_{D^*}$ for which
$D\bar{D}$ decays are forbidden: the $\psi_2$ and $\eta_{c2}$.
Of these, the $\psi_2$ is expected to have an appreciable
branching fraction for $\pipi\jpsi$ decays, making it
a preferred assignment for the $X(3872)$.  However,
this assignment has some problems:

\paragraph{Mass}~~In the charmonium picture, the $\psi_2$ mass
differs from that of its multiplet partner, the $\psi''$ with
$M=3770$~MeV, by spin-orbit and tensor interactions plus
coupled channel effects involving virtual $D\bar{D}^{(*)}$
states.  The authors of Ref.~[9] examined these effects 
and found a splitting of 66~MeV, well below
the $X(3872)$-$\psi''$ mass difference: $\Delta M=102$~MeV.

\paragraph{$\gamma\chi_{c1}$ partial width}~~The transition
$\psi_2\rt\gamma\chi_{c1}$ is an allowed $E1$ transition
with a partial width that is calculated in a potential model
(for $M_{\psi_2}=3872$~MeV) to be\cite{barnes}
$\Gamma(X\rt \gamma\chi_{c1})\simeq 360$~keV.  The
inclusion of coupled-channel effects reduces this
to\cite{elq_2} $\simeq 210$~keV.  
The Wigner-Eckart theorem says that 
the widths for $\psi_{2} \rt \pipi \jp$ and $\psi_3 \rt \pipi \jp$ should 
both be equal to $\Gamma (\psi (3770) \rt \pipi \jp)$.
The latter has been recently measured by BESII\cite{BES} and CLEO-c\cite{cleoc}
to be $80 \pm 38$~keV and $\le 55$ keV (90$\%$ CL), respectively.
These results are in some contradiction with each other, 
but it is probably
safe to say that $\Gamma (\psi (3770) \rt \pipi \jp) < 130 $ keV.
Thus, for the $X(3872) = \psi_2$ assignment, we can expect
that $\Gamma(X\rt \gamma\chi_{c1})/\Gamma(X\rt \pipi\jpsi)>1.6 $,
in contradiction with Belle's 90\% CL upper limit of 0.89.
Belle continues to search for for this decay mode
with higher sensitivity. 

\vspace{0.12in}

\subsection{\boldmath $X(3872) = h_c '$?}
The experimental situation for
the $h_c$ is unsettled, but its mass is expected to be reasonably
close to the center-of-gravity of its spin-triplet
partners, the $\chi_{c0,1,2}$ states, and safely
removed from 3872~MeV.  It has been proposed\cite{mahiko}
that the $X(3872)$ might be its first radial excitation,
the $h_c'$.  If this were the case, the $h_c'$ has
been discovered before its ground-state partner
the $h_c$;  stranger things have happened.

\paragraph{Mass}~~The $h_c '$ cannot decay to $D\bar{D}$ and
might be narrow if its mass were $3872$~MeV.  However,
even with coupled-channel effects included,  
the $h_c '$ is expected to have a mass of $\sim 3950$~MeV,
above the $D\bar{D^*}$ threshold and far from 3872~MeV.

\begin{figure}
\centerline{\psfig{file=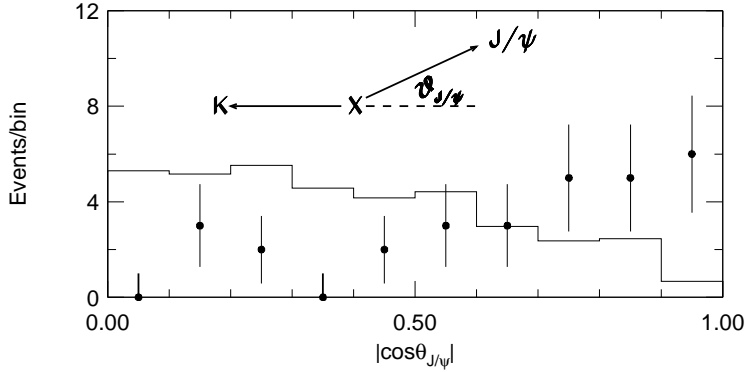,width=10cm}}
\vspace*{8pt}
\caption{
The $|\cos\theta_{\jp}|$ distribution for
$X(3872)\rt \pipi\jpsi$ events.  The histogram
shows MC expectations for a $1^{+-}$ hypothesis.
\label{fig:helicity1+-}}
\end{figure}

\paragraph{Angular distribution}~~~In Belle, the
$X(3872)$ is produced via $B\rt K X$ decays.  Since
both the initial state $B$ and the accompanying $K$ 
mesons are spin zero, the angular properties of
the final state are rather simple.\cite{mahiko}
We define $\theta_{\jp}$  as the angle between the $\jp$ and 
the negative of the $K$ momentum vectors in the $X(3872)$ rest frame
(see Fig.~\ref{fig:helicity1+-}).   
The $|\cos\theta_{\jp}|$ distribution for
$X(3872)$ events with $m_{\pipi}>$ 0.65 GeV is shown as
data points in Fig.~\ref{fig:helicity1+-}. 
For the case where the X has $J^{PC}=1^{+-}$, 
the $\cos \theta_{\jp}$ distribution would have 
a $\sin^2 \theta_{\jpsi}$ dependence.
The distribution for a MC sample of events generated
with  $1^{+-}$ expectations plus (a small) sideband-determined 
background is shown in the figure as a histogram.  
The data tend to peak near $\cos \theta_{\jp} = 1$, 
where the $1^{+-}$ expectation is zero; the $\chi^{2}/dof$ 
is very poor at 75/9, and enables us to rule out any 
$1^{+-}$ assignment for the $X(3872)$ (including the $h_c'$)  
with high confidence.

\subsection{\boldmath $X(3872) = \psi_3$?}

The authors of Refs.~[8] and [9] have suggested
that the $X(3872)$ may be the $\psi_3$, even though
$\psi_3\rt D\bar{D}$ decays are allowed and the decay 
$B\rt K\psi_3$ is likely highly suppressed. They argue 
that although $D\bar{D}$ decays are allowed, an $L=3$ 
angular momentum barrier may suppress them to such an 
extent that the $\psi_3$ might appear to be ``narrow.''

\paragraph{Width}~~Predictions (from Ref.~8)
for the total width and the
rate for $D\bar{D}$ decays are above the Belle upper 
limits, but these calculations are probably not
very reliable. 

\paragraph{$\gamma\chi_{c2}$ partial width}~~The transition
$\psi_3\rt\gamma\chi_{c2}$ is a favored $E1$ transition
with a partial width that is calculated to be $\sim$300~keV, 
where suppression due to coupled-channel effects has been
included.   Thus, the partial width for 
$\psi_{3} \rt \gamma \chi_{c2}$ is expected to be more than
twice that for $\psi_{3} \rt \pipi \jp$.

\begin{figure}
\centerline{\psfig{file=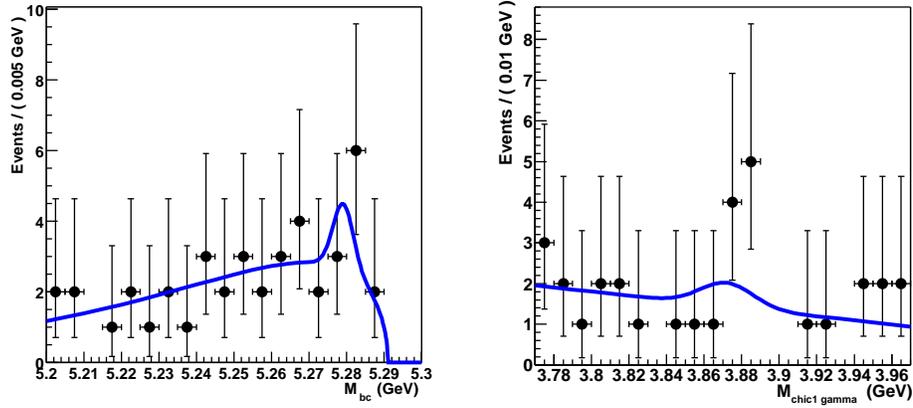,width=13cm}}
\vspace*{8pt}
\caption{
Signal-band projections of $B$-meson mass (left) and 
$M_{\gamma \chi_{c2}}$ 
(right) distributions
for events in the X(3872) region with the 
results of the unbinned fit superimposed.
\label{fig:x2gammachic2}}
\end{figure}

Belle searched for $X \rt \gamma \chi_{c2}$ using a procedure that
closely follows that used for the $\gamma \chi_{c1}$ limit 
reported in Ref.~[1].   In a sample of $B\rt K\gamma\gamma\jpsi$
event candidates, one of the $\gamma \jp$ combinations was
required to be within $\pm 10$~MeV of the $\chi_{c2}$ mass, 
and the $\gamma\chi_{c2}$
mass was required to be within $\pm 20$~MeV of  the $X(3872)$ mass.  
The results of an unbinned two-dimensional likelihood fit
to the $B$-meson mass\footnote{We use the ''beam constrained''
mass: $M_{bc}=\sqrt{(E_{cm}/2)^2-|\vec{p}_B|^2}$, where $\vec{p}_B$ is the
candidate $B$ meson's momentum in the center of mass frame.}
($M_{bc}$) and the $\gamma\chi_{c2}$ mass distributions are shown in
Fig.~\ref{fig:x2gammachic2}.
There is no evidence for a signal; the fitted signal yield
is  $2.9 \pm 3.0 \pm 1.5$ events, where the first error is statistical and
the  second systematic. The latter is estimated by the changes that occur
when  the input parameters to the fit are varied over their allowed range
of values.  This yield translates  into a 90\% CL upper limit of
$\Gamma(X\rt \gamma\chi_{c2})/\Gamma(X\rt \pipi\jpsi) < 1.1, $
below expectations for the $\psi_3$.

\begin{figure}
\centerline{\psfig{file=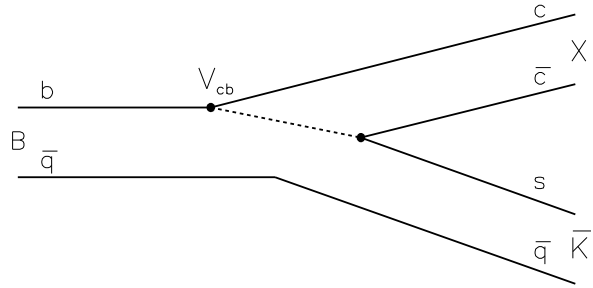,width=8.5cm}}
\vspace*{8pt}
\caption{
The leading diagram for $B\rt K(c\bar{c})$ decays.
\label{fig:b_2_kccbar}}
\end{figure}

\paragraph{Spin} ~~
$B$-meson decays to a kaon plus  a $c\bar{c}$ 
pair are expected to proceed via the
diagram shown in Fig.~\ref{fig:b_2_kccbar}.  In the 
spectator picture, the $c\bar{c}s$ quark system 
has the same spin as the
decaying $b$-quark ($i.e.$ $J=\frac{1}{2}$).  This implies $J=0$~or~1 for the
$c\bar{c}$ system.  Higher values of $J$ can be 
accomplished via the exchange of hard gluons between the
$c\bar{c}$ quarks and the ''spectator'' $\bar{q}$, but this is
expected to be suppressed. 
Thus, it is expected that for $B\rt K (c\bar{c})$
decays, $J=0$ and $J=1$ $c\bar{c}$ systems should
dominate.  The branching fractions for $B\rt K\eta_c$,
$K\jpsi$ and $K\chi_{c1}$  (with $J_{c\bar{c}}=0,~1$ and 1, 
respectively) are all about the
same (${\cal B}\simeq 10^{-3}$); in contrast, the 
decay $B\rt K \chi_{c2}$ (where $J_{c\bar{c}}=2$)
has yet to be observed (see Fig.~\ref{fig:x2gammajpsi}
below).   Reference~8 predicts a $\psi_3\rt D\bar{D}$ width  
of 4~MeV, more than 
an order of magnitude above that for $\pipi\jpsi$.  According to
Eq.~\ref{eq:product_br}, this would mean that for an $X(3872)=\psi_3$ 
assignment, the total branching fraction for $B\rt K\psi_3$ would be 
$\simeq 10^{-3}$, comparable to
that for $B\rt K\eta_c$ or $K\jpsi$. 
This seems unlikely for a $J_{c\bar{c}} = 3$ state.

\section{\boldmath $C=+1$ assignments}

If the $X(3872)$ is a $C=+1$ state, the dipion system in the $\pipi\jpsi$ 
final states would be from a $\rho\rt\pipi$ decay.  This is supported by
the dipion mass spectrum, which is concentrated near the $\rho$-meson 
mass.  However,  charmonium states all have zero isospin and, thus, decays
to $\rho\jpsi$ are isospin violating and suppressed. 
Thus, $C=+1$ charmonium assignments seem rather implausible.  Nevertheless,
since some of these have been proposed as possibilities, we address them
here.

\subsection{\boldmath $X(3872) = \chi_{c1} ' $?}
According to the authors of Ref.~[8], if the
$\chi_{c1}'$ mass were 3972~MeV, it would
have a total width less than 2~MeV.  However
the isospin-violating decay $\chi_{c1}'\rt\pipi\jpsi$
is not expected to be prominent.   There is
one well established isospin-violating hadronic
charmonium transition: $\psip\rt\pi^0\jpsi$.  This
has a measured partial width of $\simeq 0.3$~keV, which
is a factor of $\sim$500 smaller than the $\simeq 150$~keV
width for isospin-conserving $\psip\rt\pi\pi\jpsi$
process.

\paragraph{Mass}~~Potential models\cite{barnes} prefer a 
$\chi_{c1}'$ mass in the range $3930 \sim 3990$~MeV and the
coupled-channel corrections of Ref.~[9] tend to shift
this upwards, away from 3872~MeV.

\paragraph{$\gamma\jpsi$ partial width}~~A potential model
estimate\cite{barnes}  for the
$\gamma\jpsi$ partial width for a $\chi_{c1}'$ with a mass 
of 3872~MeV is 11~keV; coupled-channel effects 
may reduce this, but not by as much as an order of 
magnitude.\cite{elq_2}  If we assume that the partial width 
for the isospin-violating process $\chi_{c1}'\rt\pipi\jpsi$ 
is similar to $\Gamma(\psip\rt\pi^0\jpsi)$~($\simeq 0.3$~keV), 
we can expect that $\chi_{c1}'\rt\gamma\jpsi$ decays will be
more common than $\pipi\jpsi$ decays by at least an 
order-of-magnitude.

\begin{figure}
\centerline{\psfig{file=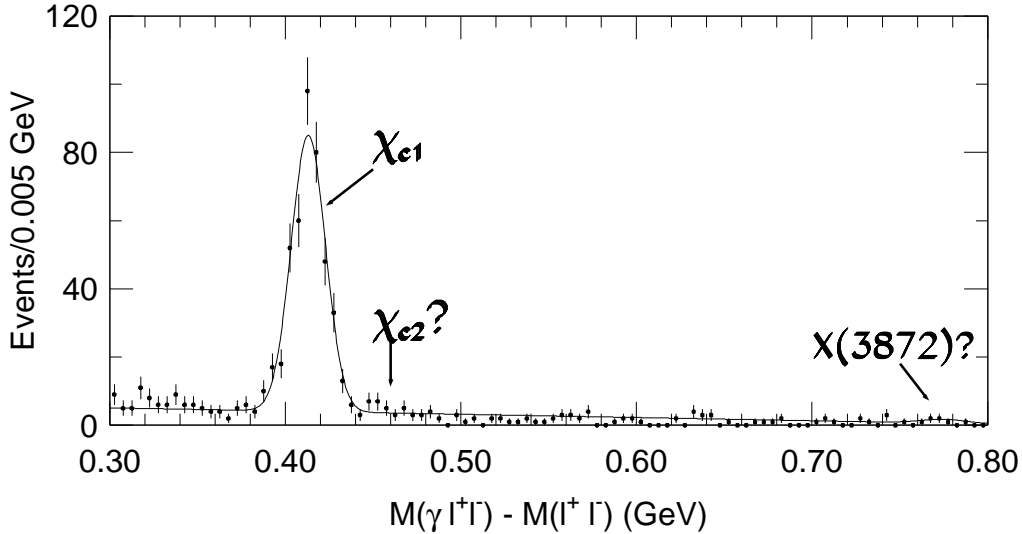,width=14cm}}
\vspace*{8pt}
\caption{
The $M_{\gamma \ell^+\ell^-}$ - $M_{\ell^+\ell^-}$
distribution for $B \rt K \gamma\jpsi$ event candidates.
The large ($\simeq470$~event) peak near $0.414$~GeV is due 
to $B\rt K\chi_{c1}$.  There are no evident signals for 
$B\rt K X(3872)$, which would show up as a peak at 
$0.776$~GeV  (or $B\rt K\chi_{c2}$,
which would peak at $0.460$~GeV).
\label{fig:x2gammajpsi}}
\end{figure}

Belle searched for $B\rt KX$; $X\rt\gamma\jpsi$ decays.
Candidate $B^{+} \rt K^{+} \gamma \jp$ events were
selected using  the criteria described in Ref.~[1].  
Figure~\ref{fig:x2gammajpsi} shows the 
$\Delta M = M_{\gamma\jpsi} - M_{\jpsi}$
distribution for selected events.  There
is a large peak at $\Delta M = 0.414$~GeV
corresponding to  $B^{+} \rt K^{+} \chi_{c1}$; 
$\chi_{c1} \rt \gamma \jp$ decays, but no sign 
of a signal at $\Delta  M = 0.776$~GeV, the position 
of the  $X(3872)$.

The $\chi_{c1}$ signal is a convenient calibration 
reaction.  We perform a three-dimensional unbinned
likelihood fit to the $M_{bc}$, $\Delta E$ and
$\Delta M$ distributions for events in the $\chi_{c1}$
region;  the fitted number of events is $470\pm 24$.
A similar fit to the events in the $X(3872)$ region
using parameters for the $M_{bc}$,
$\Delta E$ and $M_{\gamma\jpsi}$ signal functions
that are derived from the results of the $\chi_{c1}$ fit
scaled by MC-determined mass-dependent factors, gives a 
signal yield  of $7.7\pm 3.6$ events. When we include
the effects of systematic errors, this yield translates
into the limit
$$ \frac {\Gamma (X \rt \gamma \jp)}{\Gamma (X \rt \pipi \jp)}
    <0.40 ~~{\rm ( 90\% CL)},$$
which is considerably less than expectations for the 
$X(3872) = \chi_{c1}'$ assignment.

\subsection{\boldmath $X(3872) = \eta_{c2} $?}
The isospin-conserving transition $\eta_{c2}\rt\pipi\eta_c$
is expected to be much more common than the isospin-violating
$\pipi\jpsi$ decay.   Thus, from Eqn.~\ref{eq:product_br}
we infer that if the $X(3872)$ is the $\eta_{c2}$, 
the branching fraction ${\cal B}(X\rt\pipi\jpsi)$ would
be very small, of order 1\% or less, and the exclusive
$B\rt K X(3872)$ to a $c\bar{c}$ state with $J=2$, would 
be comparable to, or even larger than, the branching fractions 
for the angular momentum-favored decays $B\rt K\eta_c$
and $B\rt K \jpsi$.  

For this assignment,  $X(3872)\rt \pipi\eta_c$ decays
should be observable.  Belle plans to report a result
on this channel this summer.

\subsection{\boldmath $X(3872) = \eta_c '' $?}
The final candidate considered here is the $\eta_c''$.  If
the $\eta_c''$ mass were 3872~MeV,  its dominant decay
would be into two gluons.  The $\pipi\jpsi$ decay
would violate isospin and be suppressed.

\paragraph{Mass}~~In 2002, Belle\cite{skchoi_2} observed the 
$\eta_c'$ in exclusive $B\rt K K_sK\pi$ decays.  This observation 
was subsequently confirmed by CLEO\cite{cleo_etac}, 
BaBar\cite{babar_etac} and Belle\cite{pakhlov}.  The average mass 
and width values are\cite{elq_2} $M_{\eta_c'}= 3638\pm 4$~MeV and 
$\Gamma_{\eta_c'}= 19\pm 10$~MeV.  The $\psip$-$\eta_c'$ mass 
splitting is $\simeq 48$~MeV, much smaller than the 117~MeV 
ground-state $\jpsi$-$\eta_c$ splitting; this decrease in 
splitting with increasing radial quantum number is expected 
in QCD-inspired potential models.  Thus, one can reasonably 
expect that the $\psi(3S)$-$\eta_c''$ mass splitting 
will be less than 48~MeV.  Since the
mass of the $\psi(3S)$ is\cite{PDG} $4040\pm 10$~MeV, the $\eta_c''$
mass must be far above 3872~MeV.

\paragraph{Width}~~The dominant $\eta_c$ decay channel is
via two gluons and the world-average\cite{PDG} width of the $\eta_c$ is 
$17\pm 3$~MeV.  It is expected that the $\eta_c'$, which also
predominantly decays via two gluons, will be similar to
that for the $\eta_c$.  Existing measurements, while not
conclusive, are consistent with this conjecture.  It
is, therefore, reasonable  to expect that an $\eta_c''$
with mass below the $D\bar{D^*}$ threshold would have
a total width similar to that of the $\eta_c$ and larger
than the 2.3~MeV upper limit on the $X(3872)$ width.

\section{Summary of possible charmonium assignments} 
Although our knowledge of the $X(3872)$ properties
is still rather meager, none of the examined charmonium 
assignments naturally match the little we know about them.  A summary of the
discussion in the previous discussion is provided
in Table~\ref{tbl:summary}.

\begin{table}[h]
\tbl{\label{tbl:summary} Status of the candidate charmonium assignments.}
{\begin{tabular}{@{}cccc@{}} \toprule
State     & nickname     & $J^{PC}$ &            comment                                      \\ \colrule
$1^3D_2$  & $\psi_2$     & $2^{--}$ &   Mass wrong;  $\Gamma_{\gamma\chi_{c1}}$ too small        \\
$2^1P_1$  &  $h_c'$      & $1^{+-}$ &    Ruled out by $|\cos\theta_{\jpsi}|$ distribution     \\
$1^3D_3$  & $\psi_3$     & $3^{--}$ & $\Gamma_{\gamma\chi_{c2}}$ too small;
                                                    spin seems too high                          
\\
$2^3p_1$  & $\chi_{c1}'$ & $1^{++}$ &           $\Gamma_{\gamma\jpsi}$ too small               \\
$1^1D_2$  & $\eta_{c2}$  & $2^{-+}$ &  ${\cal B}(\pipi\jpsi)$ expected to be very small       \\
$3^1S_0$  & $\eta_c''$   & $0^{-+}$ &              Mass and width are wrong                   \\ \botrule

\end{tabular}}
\end{table}
\noindent
We conclude that if the $X(3872)$ is in fact a charmonium state, the standard
quarkonium theory needs some considerable improvements.

\section{Non-charmonium Possibilities}

The absence of an obvious charmonium assignment naturally leads one to speculate about 
non-charmonium possibilities.  In light of the close proximity of
the $X(3872)$ to $M_{D^0} + M_{D^{*0}}$~($=3871.5 \pm 1.0 $~MeV),\cite{PDG} an obvious 
candidate is a $D\bar{D^*}$ molecule-like bound state, and idea that has been
around for some time\cite{molecule_old} and has recently been
resurrected.\cite{voloshin,mahiko,tornqvist,close-1,molecule_new}  An inter-mesonic force
mediated by single pion exchange would be attractive for
$J^{PC} = 1^{++}$ or $0^{-+}$.\cite{tornqvist}  This can be
checked by measuring the
$J^{PC}$ of the $X(3872)$, which can be done by a full angular
analysis of the $\pipi\jpsi$ system.  However, this will require
more data than are currently available.

Another, perhaps less likely, possibility is some kind of a 
$c\bar{c}g$ hybrid state.\cite{close-1,hybrid}  These
are found in lattice QCD, but generally with masses of 4400~MeV or so.

\section{Conclusion}

The $X(3872)$ particle is proving to be an interesting experimental
and theoretical puzzle and a case where experiment appears to be
way ahead of theory.  As an experimentalist, the fun part about
working on this is that I have no idea where it will lead.

\section*{Acknowledgments}

I thank the organizers for inviting me to present these results
and for the excellent and well managed organization of the MESON 2004
Workshop in the interesting city of Krakow.
I especially thank my Belle colleagues from the Krakow Institute
of Nuclear Physics for their generous hospitality.  I regret that
when I finally got to visit this beautiful city, my 
friend and collaborator
Krzysztof Rybicki is no longer with us.

\appendix

\end{document}